\documentclass[doublecol]{epl2}

\title{Chopped nonlinear magneto-optic rotation:
a technique for precision measurements}
\shorttitle{Chopped nonlinear magneto-optic rotation}

\author{Harish Ravishankar \and Sapam Ranjita Chanu \and Vasant Natarajan \thanks{E-mail: \email{vasant@physics.iisc.ernet.in}}}
\shortauthor{Harish Ravishankar \etal}

\institute{Department of Physics, Indian Institute of
Science, Bangalore 560\,012, INDIA}


\abstract{
We have developed a technique for precise measurement of small magnetic fields using nonlinear magneto-optic rotation (NMOR). The technique relies on the resonant laser beam being chopped. During the on time, the atoms are optically pumped into an aligned ground state ($\Delta m=2$ coherence). During the off time, they freely precess around the magnetic field at the Larmor frequency. If the on-off modulation frequency matches (twice) the Larmor precession frequency, the rotation is resonantly enhanced in every cycle, thereby making the process like a repeated Ramsey measurement of the Larmor frequency. We study chopped-NMOR in a paraffin-coated Cs vapor cell. The out-of-phase demodulated rotation shows a Lorentzian peak of linewidth 85~$\mu$G, corresponding to a sensitivity of $0.15$~nG/$\sqrt{{\rm Hz}}$. We discuss the potential of this technique for the measurement of an atomic electric-dipole moment.
}

\begin{document}
\bibliographystyle{epl}
\maketitle

The well-known phenomenon of the rotation of the plane of
polarization of near-resonant light passing through an
atomic vapor in the presence of a longitudinal magnetic
field, called magneto-optic rotation (MOR) or the Faraday effect, is understood
to arise from the birefringence induced by the magnetic
field. This phenomenon of MOR shows nonlinear effects when
the light is sufficiently strong, as can be produced by a
laser. In the simplest manifestation of nonlinear MOR
(NMOR), the strong light field aligns the atom by inducing (through optical pumping)
$\Delta m = 2$ coherences among the magnetic sublevels. The
aligned atom undergoes Larmor precession about the
longitudinal field, and causes additional rotation due to
the precessed birefringence axis. The effect is nonlinear
because the degree of optical alignment is dependent on the
light intensity. NMOR in atomic vapor allows optical
detection of zero-field level crossings in the shot noise
limit, and has important applications in sensitive
magnetometry \cite{BKR00a,ROB07}, search for a permanent
electric dipole moment (EDM) \cite{KBE01}, and magnetic
resonance imaging (MRI) \cite{BGK02}.

In this work, we demonstrate a technique for measuring
NMOR which is ideally suited for precision measurements. The basic idea is similar to a repeated Ramsey
separated-oscillatory-fields \cite{RAM56} measurement of
the Larmor precession frequency. This is achieved by chopping the laser beam on and off. During its on time, the atoms
are optically pumped into an aligned $\Delta m=2$ coherence of the ground state.
During the off time, they (freely) precess around the
magnetic field at the Larmor frequency. If the on-off
modulation frequency matches the Larmor precession
frequency, the atoms are realigned exactly when the light
field comes back on, thus resonantly enhancing the optical
pumping process. The process continues again for the next
on-off cycle, so this is like a repeated Ramsey method.
Demodulation of the rotation at the on-off modulation
frequency shows a narrow peak at the value of the field
where the modulation frequency is equal to $2\times$ the Larmor frequency. The factor of 2 appears because the atomic alignment ($\Delta m =2$ coherence) has two-fold symmetry.

As expected, the resonance width is limited by the
decoherence time of the atomic alignment. Since the
mean-free-path in a vapor cell at room temperature is
several orders-of-magnitude larger than the cell size, the
coherence can be destroyed by spin-exchange collisions with
the walls. It has been shown that the use of paraffin
coating on the walls reduces the ground-state
depolarization rate and enhances the optical pumping
signals \cite{BOB66}. We therefore use a paraffin-coated
vapor cell for our experiments. We study the rotation on
the $D_2$ line of Cs and observe linewidths of 85~$\mu$G. This width is smaller than the width of 136~$\mu$G seen for the zero-field resonance of normal NMOR, and shows that our technique gives a precise way of measuring the Larmor precession frequency.

A previous study of Ramsey fringes in NMOR had used a thermal beam of Rb atoms with separated pump and probe regions
\cite{SKW93}. A strong pump was used to cause the alignment
and a weak probe to measure the rotation, with a separation
of 5~cm between the two regions. The resulting line shape
was dispersive and had a width of about 6~mG. By
contrast, we use the same beam for both pumping and
probing. Still the rotation appears with high
signal-to-noise ratio. The demodulated signal has the same dispersive lineshape as normal NMOR for the in-phase quadrature, but a near-perfect Lorentzian line shape for the out-of-phase quadrature. Furthermore, our linewidth of 85~$\mu$G is almost a factor of 100 smaller than the previous work. These advantages will be useful in almost all of
the applications of NMOR, particularly in the search for an
electric dipole moment (EDM).

Earlier work on NMOR has shown that modulation techniques, both frequency modulation (FM-NMOR \cite{BKY02}) and amplitude modulation (AMOR \cite{GKP06}), results in the kind of non-zero field resonances that we observe. However, our simpler technique of chopping has several advantages. In contrast to FM-NMOR, we can operate at higher frequencies, we do not have phase delays, and we can lock the laser on resonance and average the resulting rotation. And, in comparison to AMOR, our technique gives Lorentzian peaks for the out-of-phase component while AMOR results in highly non-Lorentzian lineshapes \cite{GKP06}.

\begin{figure}
\centering{\resizebox{0.95\columnwidth}{!}{\includegraphics{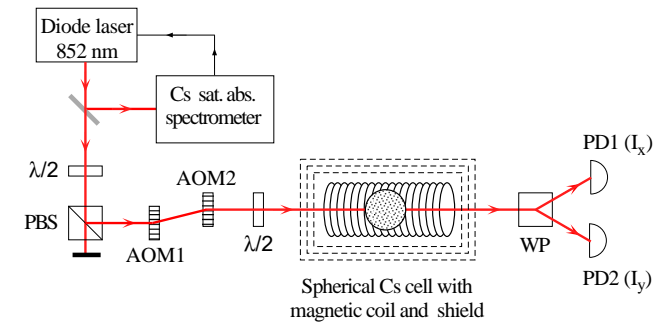}}}
\caption{(Color online) Schematic of the experiment. Figure
labels: $\lambda/2$, halfwave retardation plate; PBS,
polarizing beam splitter; AOM, acousto-optic modulator; WP, Wollaston prism; PD, photodiode}
 \label{schema}
\end{figure}

The experimental schematic is shown in Fig.\ \ref{schema}.
The NMOR measurements were done in a 80-mm diameter spherical
cell with paraffin coating on the wall. The cell was kept inside a three-layer
magnetic shield that reduced the ambient fields to below
0.1~mG. The cell had an axial solenoid coil of dimensions
140-mm diameter $\times$ 360-mm length which could produce fields
up to 1~G, calibrated using a three-axis fluxgate
magnetometer with 1~mG resolution. The field was measured
to be uniform to better than 1\% over the cell size. The field was swept linearly using a function generator.

The primary laser beam was derived from a home-built
grating-stabilized diode laser with a linewidth of 1~MHz
\cite{BRW01}. It was locked to the $F=4 \rightarrow F'=5$
closed transition of the $D_2$ line (${6S}_{1/2}
\rightarrow {6P}_{3/2}$ transition) of $^{133}$Cs using
saturated-absorption spectroscopy in a vapor cell. The
laser beam in the experimental cell had a diameter of about
2~mm and power of 20~$\mu$W. The input was linearly polarized using a polarizing
beamsplitter cube (PBS). It was passed through an
acousto-optic modulator (AOM) to allow for fast switching. A second AOM was used to compensate for the
frequency shift of the first one. After passing through the
cell, the beam was split into its linear components using a Wollaston prism. The two intensities, $I_x$ and $I_y$, were
made equal in the absence of a field (by adjusting the
halfwave retardation plate in front of the cell). The
rotation in the presence of a field is given by
\begin{equation}
\phi=\frac{1}{2} \arcsin {\frac{I_{x}-I_{y}}{I_{x}+I_{y}}}
\approx \frac{I_{x}-I_{y}}{2(I_{x}+I_{y})} \, ,
\end{equation}
where the approximation is valid for small rotation angles.
Since the rotation in our experiments was on the order of a few mrads, the measured difference $I_{x}-I_{y}$ was taken to be
the rotation signal.

\begin{figure}
\centering{\resizebox{0.91\columnwidth}{!}{\includegraphics{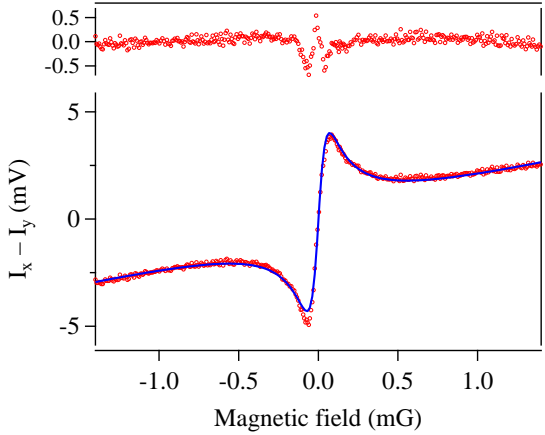}}}
\caption{(Color online) Nonlinear magneto-optic rotation (shown as the difference signal) as a function of the longitudinal magnetic field. The solid curve is a fit to Eq.\ \ref{dipersive} in the text, with the fit residuals shown on top. The fit linewidth is 136~$\mu$G.}
 \label{nmor}
\end{figure}

We first consider the phenomenon of normal NMOR. The measured rotation as a function of applied magnetic field is shown in
Fig.\ \ref{nmor}. To understand the lineshape theoretically, consider the following. The origin of the NMOR effect is the creation of a ground-state alignment ($\Delta m = 2$ coherence), and the precession of this coherence in the magnetic field. The energy shift of a magnetic sublevel ($m_F$) in a magnetic field ($B$) is $g_F \mu_B m_F B$, where $g_F$ is the Land\'e $g$ factor and $\mu_B = 1.4$~MHz/G is the Bohr magneton. With the Larmor precession frequency defined as $\omega_L = g_F \mu_B B / \hbar$, the precession frequency of the alignment is simply $2\omega_L$. If the alignment is assumed to decohere exponentially at a rate $\Gamma_r/2$, then the time-averaged rotation angle can be derived as
\begin{equation}\label{dipersive}
\Phi = C \frac{2\omega_L }{(2\omega_L)^2+\Gamma_r^2/4} \, ,
\end{equation}
where $C$ is a proportionality constant that depends on the degree of alignment. Thus, we see that the lineshape of normal NMOR is dispersive and centered at $\omega_L=0$ (i.e., $B=0$), exactly as observed in the figure. The solid curve is a fit to Eq.\ \ref{dipersive} along with a linear term to account for the MOR effect. The fit yields a width of 136~$\mu$G.

\begin{figure}
(a)\centering{\resizebox{0.95\columnwidth}{!}{\includegraphics{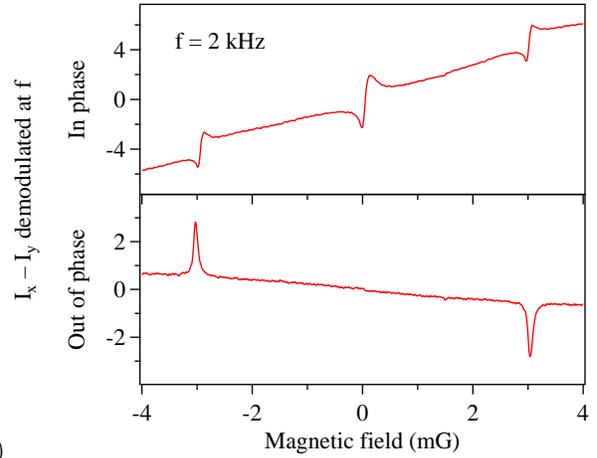}}}
(b)\centering{\resizebox{0.91\columnwidth}{!}{\includegraphics{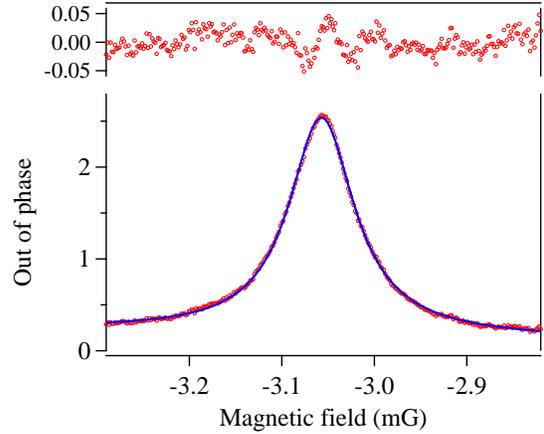}}}
\caption{(Color online) (a) NMOR demodulated at the chopping frequency $f$ plotted as a function of the longitudinal magnetic field. Both in-phase and out-of-phase components are shown. (b) A close-up of the left Lorentzian peak in the out-of-phase component. The solid curve is a Lorentzian fit (residuals on top) yielding a linewidth of 85~$\mu$G.}
 \label{cmor}
\end{figure}

Let us now turn to the NMOR signals obtained with the chopped-NMOR technique, shown in Fig.\ \ref{cmor}(a). To a rough approximation, one can say that the effect of the chopping is to modulate the atomic alignment at this frequency. Thus, if the chopping frequency is $\omega_m$, the constant $C$ in Eq.\ \ref{dipersive} becomes $C \cos(\omega_m t)$. The lock-in amplifier demodulates this signal at $\omega_m$ to get the in-phase and out-of-phase quadratures. It is straightforward to see that the time-averaged rotation angle for the in-phase component is
\begin{equation}
 \label{inphase}
\Phi_{\rm ip} = C \left[ \frac{D(\omega_L)}{2} + \frac{D(\omega_L - \omega_m)}{4} + \frac{D(\omega_L + \omega_m)}{4} \right] ,
\end{equation}
where $$ D(x) = \frac{2x}{(2x)^2+\Gamma_r^2/4} $$
is the same dispersive function that appeared in Eq.\ \ref{dipersive}. That is, the in-phase rotation has two additional peaks compared to normal NMOR, which have the same lineshape but are centered at $\pm \omega_m$ and have half the amplitude. This is exactly what is seen in the figure.

A similar analysis shows that the out-of-phase rotation is given by
\begin{equation}
 \label{outphase}
\Phi_{\rm op} = C \left[ \frac{L(\omega_L - \omega_m)}{4} - \frac{L(\omega_L + \omega_m)}{4} \right] ,
\end{equation}
where
$$ L(x) = \frac{\Gamma_r}{(2x)^2+\Gamma_r^2/4} $$
is a Lorentzian function. As seen in the figure, the out-of-phase component is a sum of two Lorentzian peaks centered at $\pm \omega_m$. The linewidth (i.e., full width at half maximum) of each peak  is $\Gamma_r$, determined by the same decoherence rate as for normal NMOR. A fit to a Lorentzian lineshape yields a width of 85~$\mu$G. This is smaller than the near-zero-field width of 136~$\mu$G seen for normal NMOR. It has been shown that the width of the NMOR resonance is critically dependent on both the residual field and gradients \cite{PJR06}. The relative importance of these imperfections is less when we consider the Larmor precession at the large field, as in the case of the chopped NMOR. This difference could account for the smaller linewidth that we observe.

An important application of our technique is to use it for the measurement of an EDM of an atom. In such applications, it is advantageous to use Rb as a co-magnetometer in the same cell. This is because Rb has a smaller EDM enhancement factor, defined as the ratio of the atomic EDM ($d_A$) to the intrinsic electron-EDM ($d_e$). The enhancement factor is 120.53 for Cs and 25.74 for Rb \cite{NSD08}. Therefore, both atoms will be sensitive to the {\it magnetic field} in the same way but to the {\it electric field} in a different way, which gives us an experimental handle on potential systematic errors. To enable this, we have filled our cell with both Rb and Cs. The experiment now has a second laser beam, which is locked to the $F=2 \rightarrow F'=3$ transition of the $D_2$ line in $^{87}$Rb (at 780~nm). The two beams are linearly polarized in orthogonal directions and mixed on a PBS. Both beams are chopped at the same frequency and pass through a zero-order halfwave plate oriented at 45$^\circ$, so that the rotation after the cell can be measured using the same Wollaston prism and photodiodes.

\begin{figure}
\centering{\resizebox{0.95\columnwidth}{!}{\includegraphics{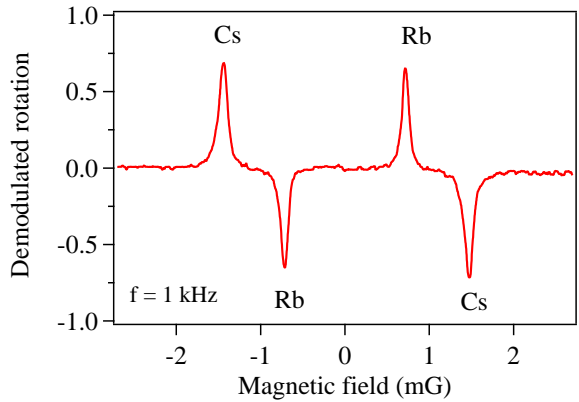}}}
\caption{(Color online) Use of Rb as a co-magnetometer. The NMOR signal demodulated at the chopping frequency $f$ plotted as a function of the longitudinal magnetic field. Only the out-of-phase component is shown. The $g$ factors of Rb and Cs differ by a factor of 2.}
 \label{comag}
\end{figure}

The out-of-phase component of the demodulated signal is shown in Fig.\ \ref{comag}. As expected, there are two sets of peaks, one for Cs and one for Rb. The peaks for Rb appear at a field that is exactly a factor of two smaller because its Land\'e $g$ factor is $2 \times$ larger. The linewidths of the peaks for Rb are also about 20\% smaller than that seen for Cs.

Since the width of the NMOR resonance depends on magnetic-field inhomogeneities \cite{PJR06}, we hope to reduce it below 10~$\mu$G with better shielding and by using compensation coils. Indeed, widths of the zero-field NMOR resonance of order 2~$\mu$G have been observed using a four-layer shield \cite{KBE01}. However, even with our width of 85~$\mu$G, the shot-noise-limited sensitivity to a measurement of the magnetic field is $\delta B = 0.15$~nG/$\sqrt{{\rm Hz}}$. The measurement of an atomic EDM consists of measuring the shift in the Larmor precession frequency in the presence of an external electric field ($E$) due to the $\vec{d_A} \cdot \vec{E}$ interaction \cite{NAT05}. Thus, the sensitivity $\delta d_A$ to an EDM is related to the smallest detectable change of the magnetic field as,
$$\delta d_A = \frac{g_F \mu_B \delta B}{2E} \, , $$
where the factor of 2 accounts for the applied electric field being parallel or anti-parallel to the magnetic field. This field reversal is built into our technique because we are sweeping the field from negative to positive values. Taking typical values of $E=10$~kV/cm and an integration time of 10 days, the shot-noise limited sensitivity to an EDM measurement is $\delta d_A = 1.2 \times 10^{-25}$~e-cm, which is an order-of-magnitude better than the best measurement in Cs of $d_{\rm Cs} < -1.8  \times 10^{-24}$~e-cm \cite{MKL89}. This implies a $125 \times$ smaller limit for $d_e$ because of the enhancement factor.

In summary, we have developed a technique to measure a small magnetic field, or equivalently the Larmor precession frequency of an atom, using the nonlinear Faraday effect. The atoms are first aligned and then probed by chopping the laser at twice the Larmor frequency. Thus, it is like a Ramsey separated-oscillatory-fields method with the pump and probe regions separated in time. The rotation is modulated at the chopping frequency and shows a near-perfect Lorentzian line shape for the out-of-phase demodulated component. The observed linewidth of 85~$\mu$G (or frequency width of 60~Hz) holds much promise for the measurement of an atomic EDM. The best limit on the electron-EDM of $1.6 \times 10^{-27}$~e-cm comes from a measurement in Tl \cite{RCS02} because of its large enhancement factor of 585. In principle, the shot-noise limited sensitivity of our technique using Cs will allow us to reach the same precision, while further improvements in the linewidth and sensitivity seem quite possible.

\begin{acknowledgments}
This work was supported by the Department of Science and
Technology, India.
\end{acknowledgments}


\end{document}